\documentclass[%
 aip,
 jmp,%
 amsmath,amssymb,
 reprint,%
]{revtex4-1}

\usepackage{graphicx}% Include figure files
\usepackage{dcolumn}% Align table columns on decimal point
\usepackage{bm}% bold math
\usepackage{multirow}
\usepackage{upgreek}
\begin{document}

\preprint{AIP/123-QED}

\title[Noble gas excimer scintillation following neutron capture in boron thin films]{Noble gas excimer scintillation following neutron capture in boron thin films}

\author{Jacob C. McComb}
\affiliation{\mbox{Nuclear Engineering Program, University of Maryland, College Park, Maryland, 20742, USA}}
\author{Michael A. Coplan}
\affiliation{\mbox{Institute for Physical Science and Technology, University of Maryland, College Park, Maryland, 20742, USA}}
\author{Mohamad al-Sheikhly}
\affiliation{\mbox{Nuclear Engineering Program, University of Maryland, College Park, Maryland, 20742, USA}}
\author{Alan K. Thompson}
\affiliation{\mbox{National Institute of Standards and Technology, Gaithersburg, Maryland, 20899, USA}}
\author{Robert E. Vest}
\affiliation{\mbox{National Institute of Standards and Technology, Gaithersburg, Maryland, 20899, USA}}
\author{Charles W. Clark}
\affiliation{\mbox{National Institute of Standards and Technology, Gaithersburg, Maryland, 20899, USA}}
\affiliation{\mbox{Institute for Physical Science and Technology, University of Maryland, College Park, Maryland, 20742, USA}}
\affiliation{Joint Quantum Institute, National Institute of Standards and Technology and University of Maryland, Gaithersburg, Maryland, 20899, USA}

\date{\today}

\begin{abstract}
Far-ultraviolet (FUV) scintillation signals have been measured in heavy noble gases (argon, krypton, xenon) following boron-neutron capture ($^{10}$B(n,$\upalpha$)$^7$Li) in $^{10}$B thin films. The observed scintillation yields are comparable to the yields from some liquid and solid neutron scintillators. At noble gas pressures of 107 kPa, the number of photons produced per neutron absorbed following irradiation of a 1200 nm thick $^{10}$B film was 14,000 for xenon, 11,000 for krypton, and 6000 for argon. The absolute scintillation yields from the experimental configuration were calculated using data from (1) experimental irradiations, (2) thin-film characterizations, (3) photomultiplier tube calibrations, and (4) photon collection modeling. Both the boron films and the photomultiplier tube were characterized at the National Institute of Standards and Technology. Monte Carlo modeling of the reaction cell provided estimates of the photon collection efficiency and the transport behavior of $^{10}$B(n,$\upalpha$)$^7$Li reaction products escaping the thin films. Scintillation yields increased with gas pressure due to increased ionization and excitation densities of the gases from the $^{10}$B(n,$\upalpha$)$^7$Li reaction products, increased frequency of three-body, excimer-forming collisions, and reduced photon emission volumes (i.e., larger solid angle) at higher pressures. Yields decreased for thicker $^{10}$B thin films due to higher average energy loss of the $^{10}$B(n,$\upalpha$)$^7$Li reaction products escaping the films. The relative standard uncertainties in the measurements were determined to lie between 14 \% and 16 \%. The observed scintillation signal demonstrates that noble gas excimer scintillation is promising for use in practical neutron detectors.
\end{abstract}

\maketitle

\section{Introduction}

Neutron detection is essential to homeland security, nuclear reactor instrumentation, neutron diffraction science, oil well logging, particle physics, radiation safety, and many other technical and commercial activities. The current shortage of $^3$He, the neutron absorber used in most gas-filled proportional counters, has created a strong incentive to develop new methods of neutron detection. Excimer-based neutron detection (END) provides an alternative with many attractive properties, including highly efficient signal, fast response time, immunity to radiation damage, unrestricted geometry, moderate gas pressure, low voltage operation, durability, and low cost of available components (e.g., no $^3$He).

\begin{figure}[h]
\centering
\includegraphics{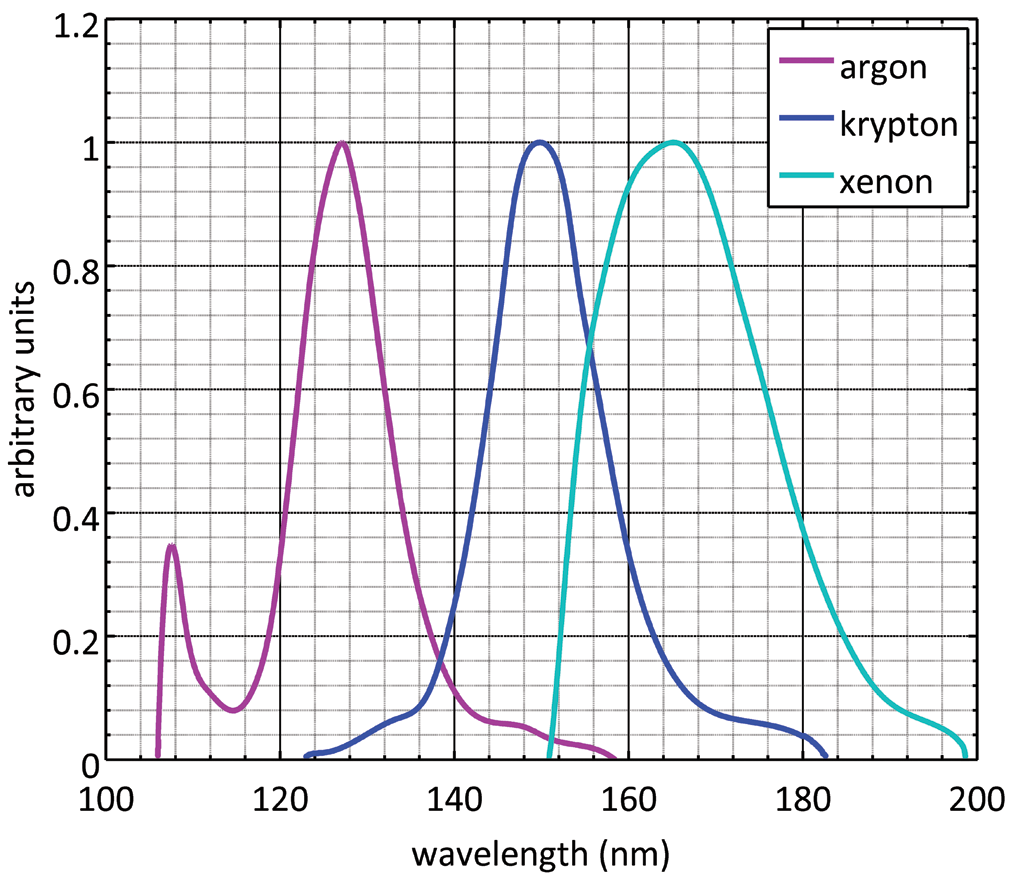}
\caption{FUV emission spectra of Ar, Kr, and Xe (left to right) excimers produced by thyratron modulator excitation. Redrawn from \cite{1}. Peak intensities have been normalized to unity.}
\label{Figure1}
\end{figure}

END relies on the same conversion mechanism as most traditional thermal neutron detectors. A neutron, when absorbed by specific nuclides (e.g., $^3$He, $^6$Li, $^{10}$B), precipitates an exothermic reaction, and the energetic charged-particle reaction products deposit their kinetic energy within a detection medium through ionization and electronic excitation. By surrounding or mixing the neutron-absorbing target with a noble gas, the charged-particle reaction products induce the formation of noble gas excimers (NGEs) as they dissipate kinetic energy. NGEs are loosely bound, diatomic molecules that exist only in an excited electronic state (e.g., Ar$_2^*$). NGEs are short-lived and decay by emitting far-ultraviolet (FUV) photons. This mechanism of photon emission, referred to as excimer scintillation, provides a unique method for detecting neutrons. Ar, Kr, and Xe produce excimers with wavelengths between 105 nm and 190 nm \cite{1,2}, as shown in the emission spectra in Figure \ref{Figure1}. A combination of features—large light output, fast decay, transparency to excimer radiation, unique decay structure, and immunity to radiation damage—make noble gases particularly suitable as radiation detection media \cite{3}.

The spectrum of the simplest excimer, He$_2^*$, was first analyzed by Hopfield \cite{4}, and the electronic structure was inferred from the positions and intensities of the emission lines. The molecular constants for the lowest excited singlet and triplet states are given in Herzberg \cite{5}. The excimers of Ne, Ar, Kr, and Xe were investigated by Tanaka et al. \cite{6,7}, and later implemented in FUV radiation sources \cite{1,8}. The use of NGE scintillation in particle detection began in the 1950s and extended into the 1980s with measurements of energy resolution and gamma sensitivity \cite{3,9,10,11,12,13}. The mechanisms of excimer formation and decay have been investigated both experimentally \cite{2,14, 15} and theoretically \cite{16,17,18}. Recently, liquid noble gas detectors have played a significant role in efforts to detect the neutrino mass and magnetic moment, and dark matter candidates, such as weakly interacting massive particles \cite{19}. Few sources provide data on the absolute scintillation yields of gas-phase noble gas scintillators \cite{20,21,22}, and the data from these sources are inconsistent. For this study, we constructed an apparatus with a well-defined geometry and carefully characterized the experimental parameters. Those parameters not readily accessible to experimental measurement were modeled with standard numerical techniques.

In a previous publication, we reported on interactions of cold neutrons with mixtures of $^3$He and heavy noble gases \cite{23}. Using CaF$_2$, sapphire, and fused silica spectral filters the resulting radiation was identified as NGE emissions. In that experiment, thousands of scintillation photons per neutron absorbed were observed. Due to the growing scarcity of $^3$He, similar results were sought using other neutron absorbing targets in the presence of heavy noble gases.

Here, we present data on NGE scintillation following the boron-neutron capture reaction ($^{10}$B(n,$\upalpha$)$^7$Li) in $^{10}$B thin films. Properties of the $^{10}$B films, the thermal neutron beam, and the photon detector package were characterized to derive the number of NGE photons per neutron absorbed. Additionally, Monte Carlo modeling was used to calculate charged-particle energy transfer and photon emission under the experimental conditions. Based on the experiments and modeling, we have determined the absolute number of NGE photons produced following the $^{10}$B(n,$\upalpha$)$^7$Li reaction for a range of noble gas pressures and $^{10}$B film thicknesses.

\section{Experiment}

\subsection{Reaction cell}

\begin{figure}[h]
\centering
\includegraphics{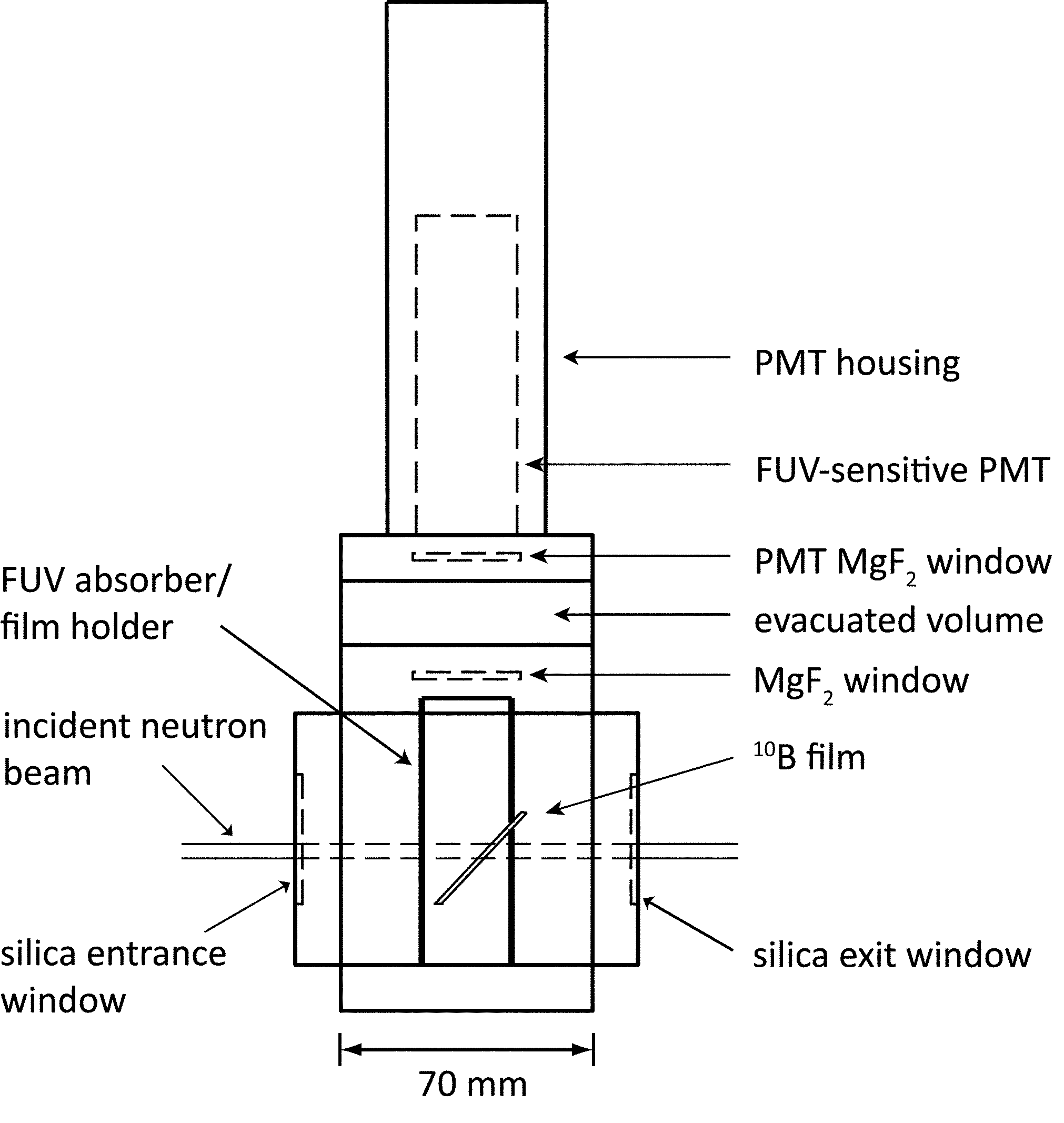}
\caption{A diagram of the reaction cell and PMT configuration.}
\label{Figure2}
\end{figure}

The reaction cell for the production and detection of excimer radiation consisted of a 70 mm stainless steel cube with metal-seal flange ports on all sides, attached to an FUV-sensitive photomultiplier tube (PMT). Silica entrance and exit windows allowed the neutron beam to pass through the cell without significant attenuation. Within the cell, a high-purity noble gas environment surrounded a $^{10}$B thin-film target held in place by a slotted aluminum cylinder beneath the PMT. The cylinder was coated with black copper oxide to reduce reflections of FUV photons. This arrangement provided a well-defined geometry for modeling the photon collection efficiency. A diagram of the apparatus appears in Figure \ref{Figure2}.

The gas-handling system connected to the reaction cell consisted of a turbomolecular pump, noble gas cylinders, a cold cathode vacuum gauge, a digital pressure gauge, a residual gas analyzer, a gas filter, a metering valve, and a series of bellows-sealed valves. The components were connected to a stainless steel manifold with welded or metal gasket fittings. Both the Kr and Xe were research grade (99.999 \% purity), while the Ar was ultra-high grade (99.9995 \% purity). The gases were passed from the manifold into the reaction cell through the filter, which removed H$_2$O, O$_2$, CO, and CO$_2$ to $< 1 \times 10^{-10}$ \%. The filter also removed acids, organics, and refractory compounds to $< 1 \times 10^{-10}$ \%, and bases to $< 5 \times 10^{-7}$ \%. Before introducing inert gas into the reaction cell, the cell was baked at 100 $^{\circ}$C overnight. The base pressure in the cell was typically $7 \times 10^{-5}$ Pa. Upon filling, the digital pressure gauge was used to determine the pressure in the reaction cell with a relative standard uncertainty of $\pm 0.1$ \% of full scale (206 kPa) due to non-linearity, hysteresis, and repeatability of the gauge.

\begin{figure*}[!t]
\centering
\includegraphics{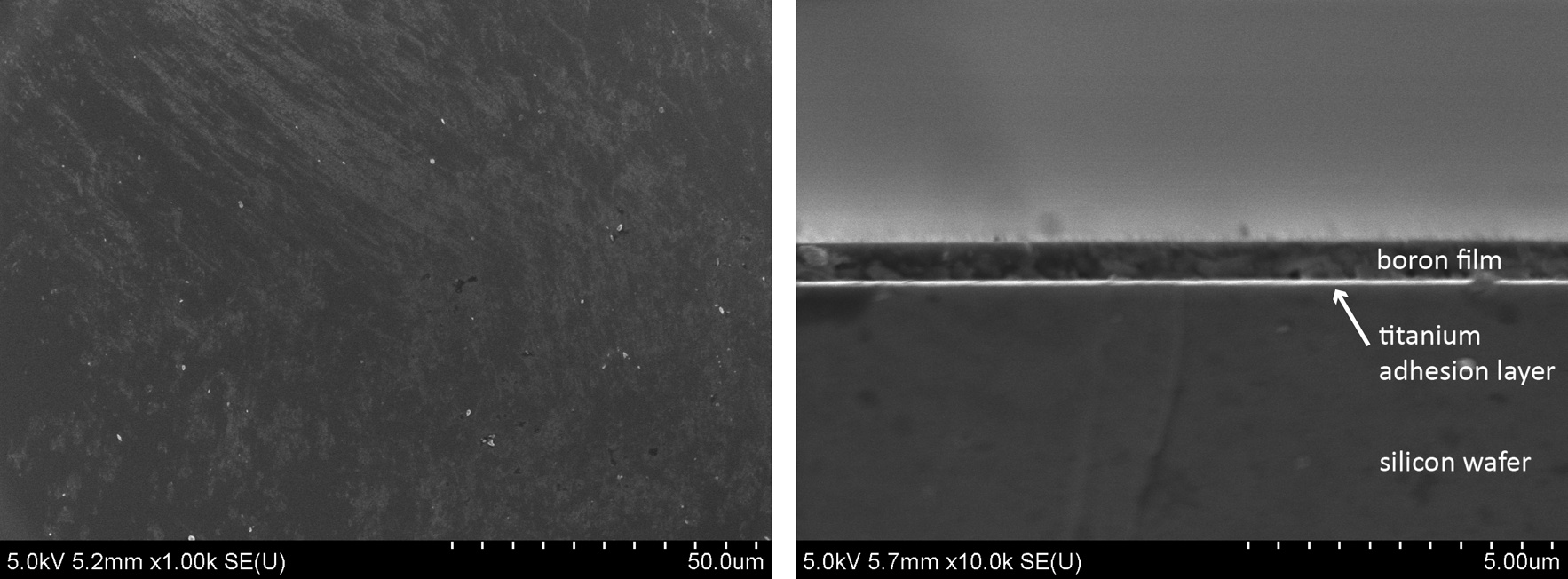}
\caption{SEM images of a boron thin film target, showing both the surface of the film at 1000$\times$ magnification (right) and a cross section of the film, adhesion layer, and substrate at 10,000$\times$ magnification (left).}
\label{Figure3}
\end{figure*}

\subsection{$^{10}$B thin films}

$^{10}$B has an isotopic abundance of 19.9 \%, and a thermal neutron absorption cross section of ($3842 \pm 8$) b \cite{24}. For nearly all thermal neutron absorptions, $^{10}$B undergoes an exothermic neutron-capture reaction, resulting in the emission of an $\upalpha$-particle and a $^7$Li ion. The combined energy of these products is either 2.79 MeV or 2.31 MeV (with branching ratios of 6 \% and 94 \%, respectively) depending on the final state of the $^7$Li nucleus. 

$^{10}$B thin-film targets were fabricated at the National Institute of Standards and Technology (NIST) Center for Nanoscale Science and Technology (CNST) by physical vapor deposition in an electron beam evaporator. The deposition source material was isotopically enriched in $^{10}$B to 92 \%. The substrates were silicon wafer pieces, 25.4 mm $\times$ 25.4 mm $\times$ 0.525 mm. To increase adhesion and reduce film stress, thin layers of titanium and chemically deposited natural boron were added to the surfaces of the silicon wafers before depositing $^{10}$B. Films were fabricated at nominal thicknesses of (300, 600, 900, and 1200) nm. 

Several characterizations of the boron thin-film targets were performed to determine their neutron absorbing properties (content, thickness, density) and the stability of those properties over the course of the scintillation measurements. These characterizations included scanning electron microscopy, x-ray photoelectron spectroscopy, neutron imaging, profilometry, and x-ray diffraction. Scanning electron microscope images of a 600 nm $^{10}$B film appear in Figure \ref{Figure3}. During the experiments, contamination of the $^{10}$B thin films from exposure to air was of particular concern. X-ray photoelectron spectroscopy scans of samples exposed to air for up to 3.5 months demonstrated no evidence of boron oxide formation within 5 nm to 10 nm of the film surface. 

\begin{figure*}[!t]
\centering
\includegraphics{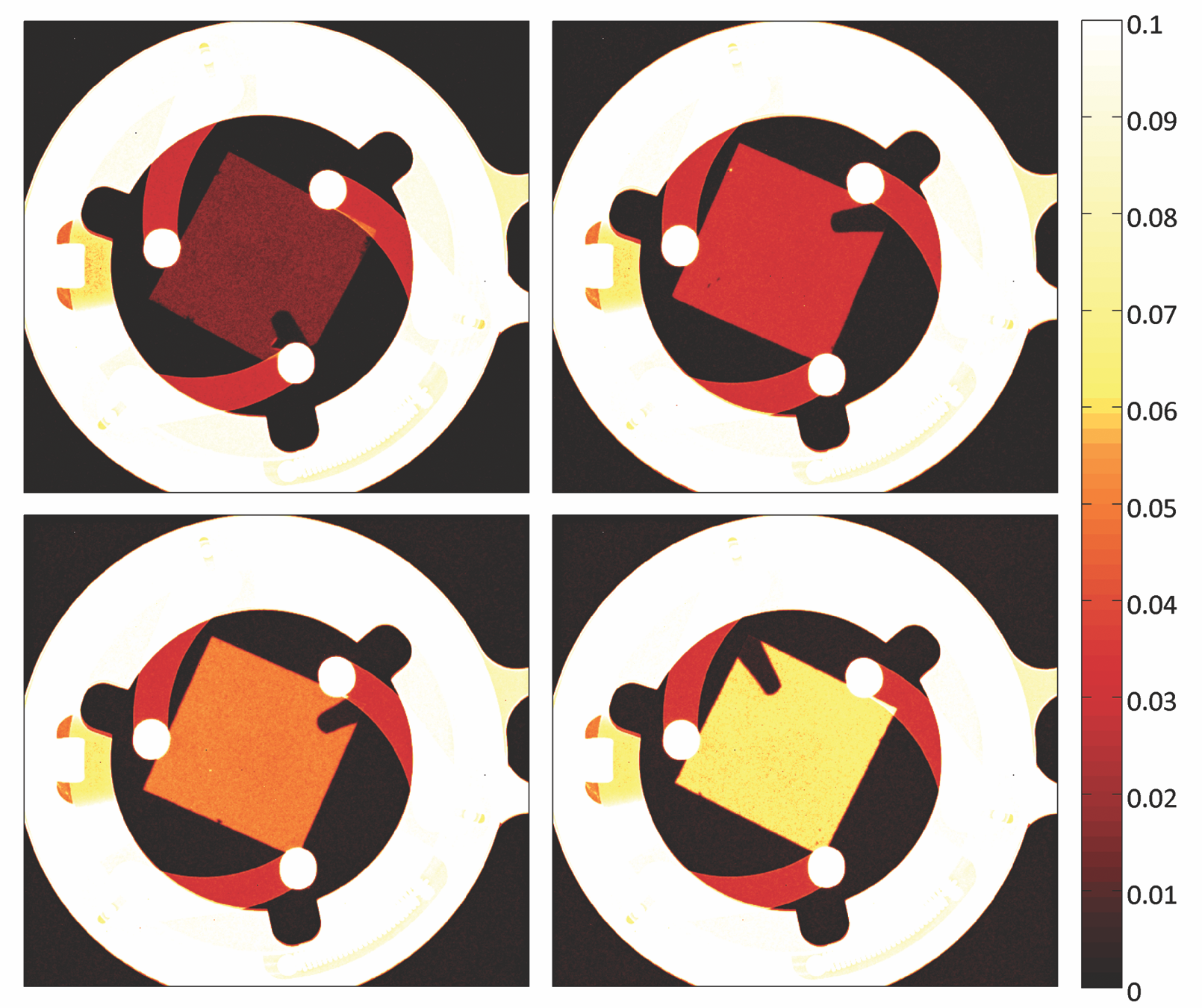}
\caption{Thermal-neutron images of the $^{10}$B thin films, as measured at the NCNR NIF. Nominal film thicknesses: 300 nm (top left), 600 nm (top right), 900 nm (bottom left), 1200 nm (bottom right). The color scale (far right) represents fractional neutron absorption between 0 \% and 10 \%.}
\label{Figure4}
\end{figure*}

Areal densities of the $^{10}$B thin films were measured at the Neutron Imaging Facility (NIF) at the NIST Center for Neutron Research (NCNR). This facility uses an intense, collimated beam of thermal neutrons to obtain radiograph images of neutron absorbing samples \cite{27}. The areal densities were used to calculate the rate of neutron capture during the NGE scintillation measurements. Radiograph images of the various film targets appear in Figure \ref{Figure4}.

\subsection{Photon detector package}

\begin{figure}[h]
\centering
\includegraphics{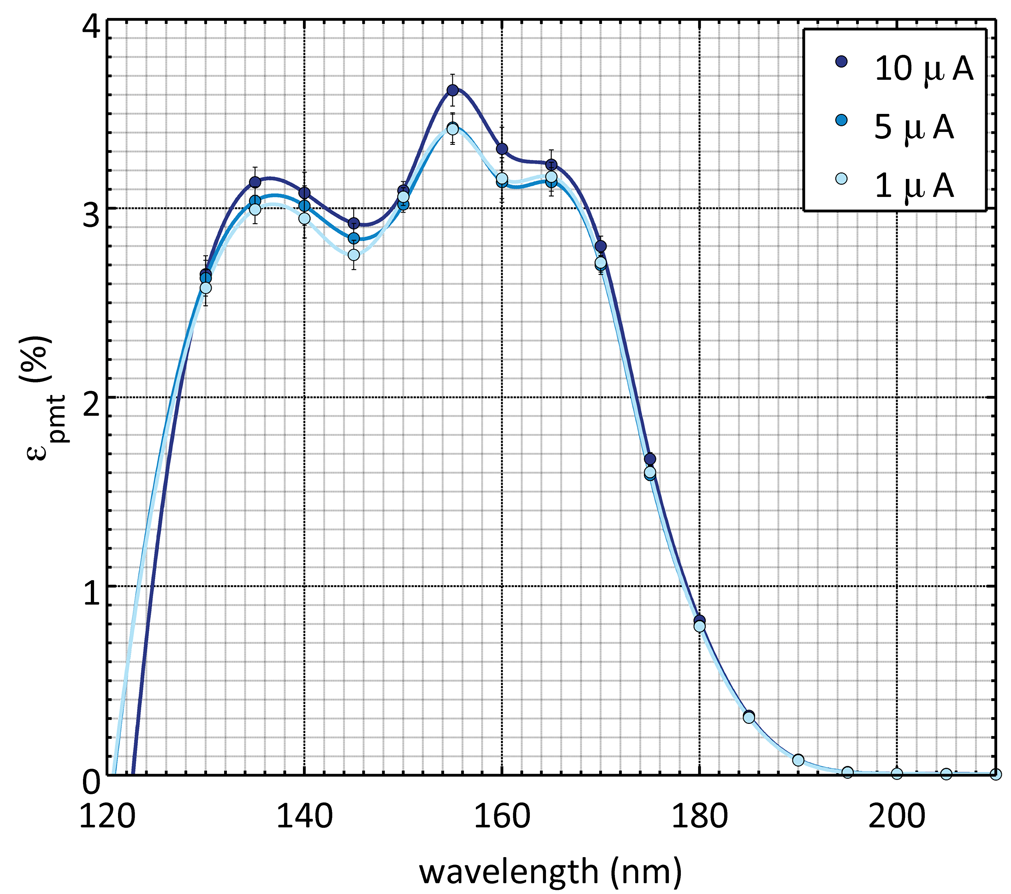}
\caption{The intrinsic efficiency of the photon detector package over the FUV region, as calibrated at SURF III. Datasets represent calibrations recorded at three distinct synchrotron electron beam currents. Cubic spline fits appear for each dataset. Error bars represent combined uncertainties from statistical uncertainties and uncertainties in the linear fits of photon flux vs.\ electron beam current.}
\label{Figure5}
\end{figure}

The photon detector package atop the reaction cell consisted of an FUV-sensitive PMT (Hamamatsu R6835 [Certain commercial equipment, instruments, or materials are identified in this paper to foster understanding. Such identification does not imply recommendation or endorsement by the National Institute of Standards and Technology, nor does it imply that the materials or equipment identified are necessarily the best available for the purpose.]) with a thin MgF$_2$ entrance window, an intermediate evacuated volume, and a second MgF$_2$ window separating the PMT from the reaction cell. The efficiency of the photon detector package was calibrated against an absolutely calibrated silicon photodiode at the Synchrotron Ultraviolet Radiation Facility (SURF III) at NIST. The continuous distribution of radiant power in the spectral region of interest (130 nm to 210 nm) from SURF III is well suited for radiometric calibrations \cite{26}. The efficiency of the photon detector package over this region appears in Figure \ref{Figure5}.  

\begin{figure}[h]
\centering
\includegraphics{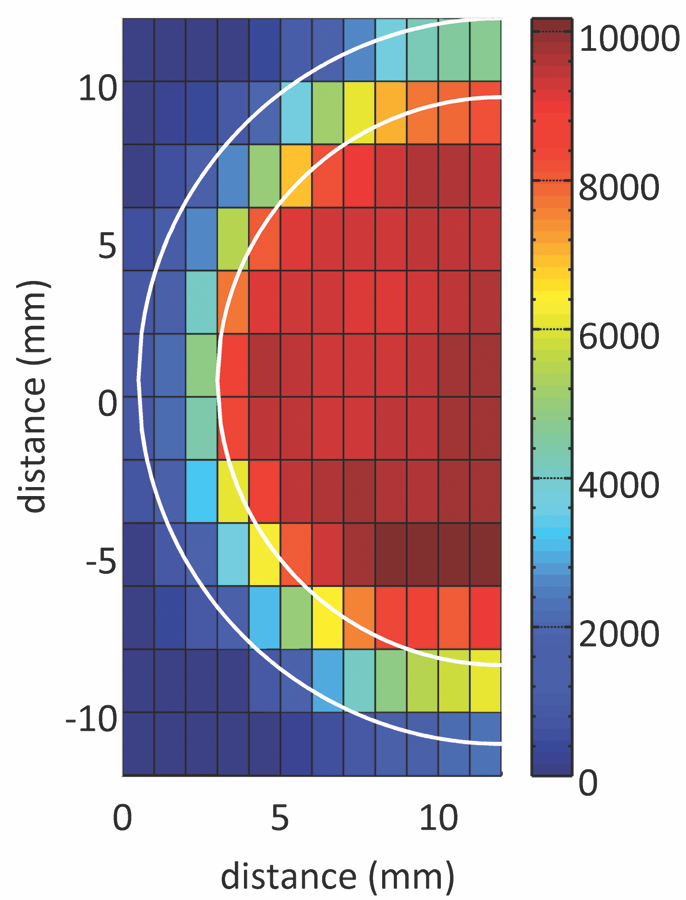}
\caption{The spatial response of photon detector package as measured at SURF III at 170 nm. Measurements were limited to approximately half of the detector area due to constraints of the experimental setup. The color scale represents photon counts over a 2 s interval, at a synchrotron electron beam current of 10 $\upmu$A. The boundaries of the tally regions used in the photon collection model appear as white semicircles with radii of 9 mm and 11.5 mm.}
\label{Figure6}
\end{figure}

The spatial response of the photon detector package was also measured at SURF III. The response was uniform (within 3 \% of the mean) over an 18 mm diameter, with a diminishing response from 18 mm to the 23 mm outer diameter of the photocathode. We attribute the diminishing response to irregularities in the MgF$_2$ windows and reduced detection efficiency for photons striking the periphery of the photocathode. A color map of the spatial response appears in Figure \ref{Figure6}. The spatial response was incorporated into the Monte Carlo model of photon collection. 

\begin{figure*}[!t]
\centering
\includegraphics{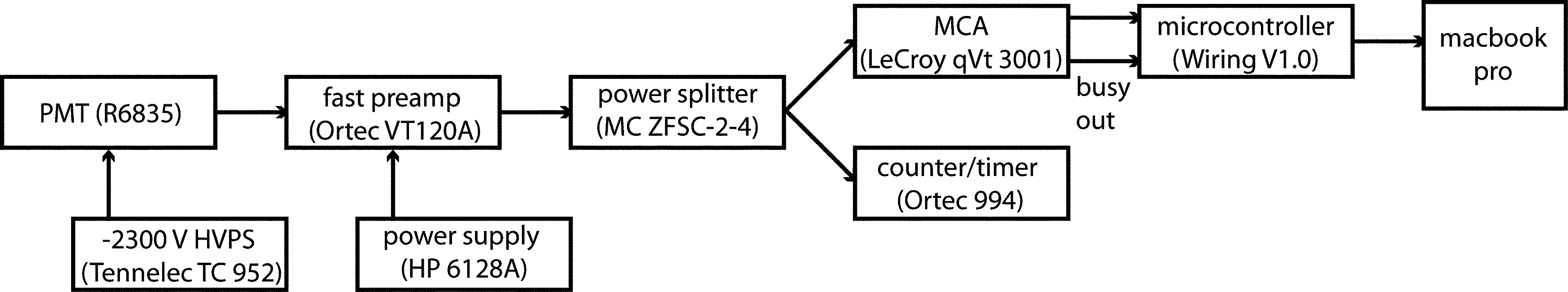}
\caption{A block diagram of the photon counting electronics. HVPS: high voltage power supply, MCA: multichannel analyzer.}
\label{Figure7}
\end{figure*}

During the NGE measurements, the PMT was operated at -2300 V, corresponding to a nominal gain of $3 \times 10^5$. Pulses from the PMT were amplified by a fast preamplifer with a gain of 200 and a rise time of $\leq 1$ ns. The amplified pulses were divided in two with a power splitter. One splitter output was sent to a counter/timer and the other to a multichannel analyzer (MCA). The MCA was used to obtain pulse-charge distributions (PCDs). These PCDs confirmed that multi-photoelectron pulses were not contributing significantly to the number of pulses counted by the counter/timer. The MCA was operated with a 20 ns integration time and a channel resolution of 0.25 pC over 1024 channels. Because the digitization time of the MCA was long with respect to the typical decay time of each scintillation event, not every pulse from the PMT was collected by the MCA. Nevertheless, the resulting PCDs were assumed to be representative of the true distributions. The counter/timer with a maximum count rate of 100 MHz and a pulse-pair resolution of $< 10$ ns recorded the true number of PMT pulses. A block diagram of the electronics appears in Figure \ref{Figure7}.

\subsection{Neutron beam}

The Maryland University Training Reactor (MUTR) provided a source of thermal neutrons for the NGE scintillation experiments. The MUTR is a TRIGA reactor with a peak power of 250 kW. The MUTR thermal column consists of 1.5 m of graphite adjacent to the core and a custom insert to produce a collimated thermal neutron beam with a diameter of 50 mm.

\begin{figure*}[t]
\centering
\includegraphics{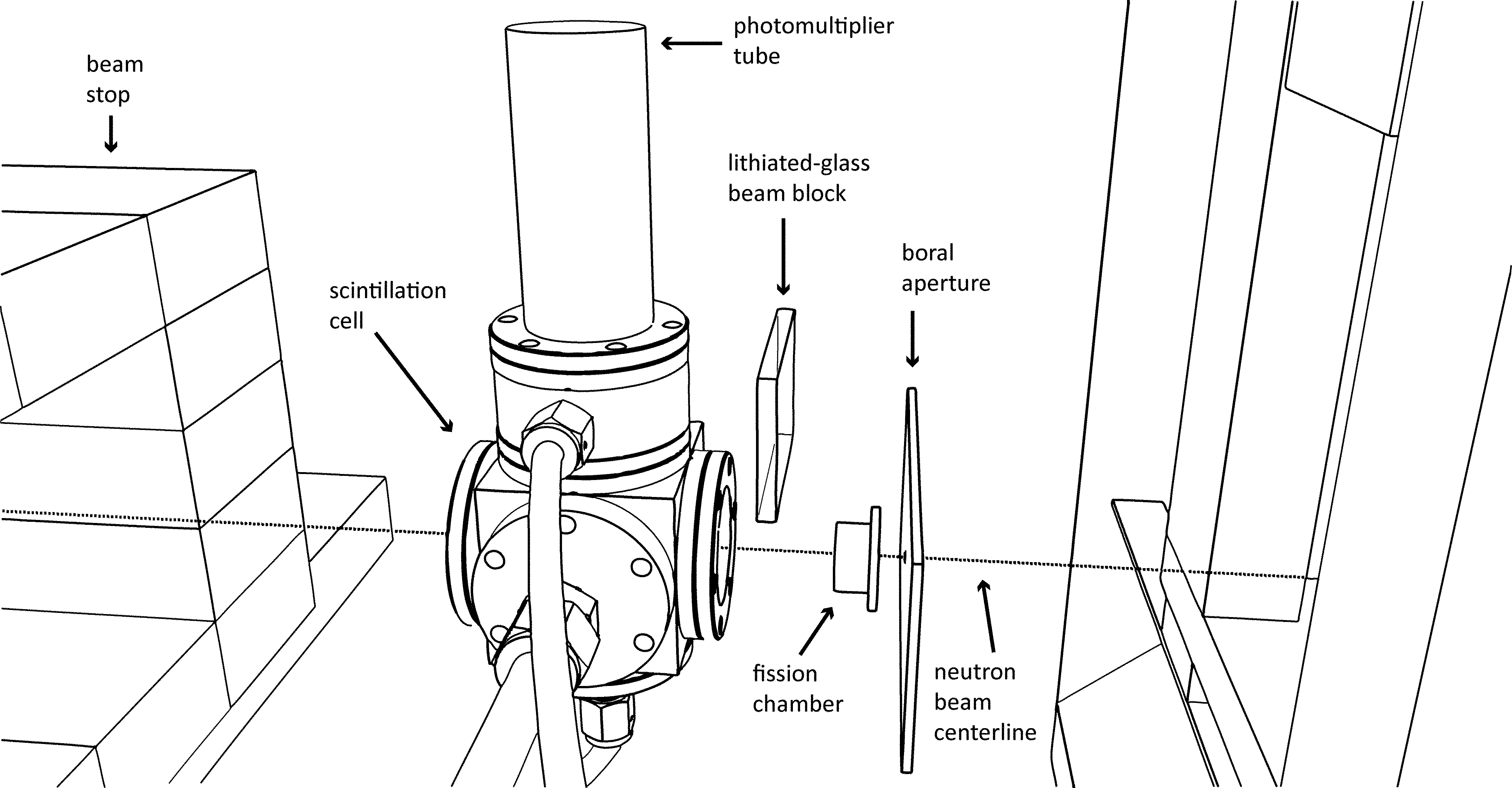}
\caption{A diagram of the excimer scintillation experiment. The collimated thermal neutron beam from the MUTR thermal column propagates from right to left.}
\label{Figure8}
\end{figure*}

Beyond the collimator, in front of the reaction cell, a borated aluminum aperture reduced the neutron beam to a 4 mm diameter. The neutron fluence behind this aperture was monitored with a NIST-calibrated fission chamber throughout the experimental irradiations. The fission chamber was specifically designed as a beam monitor, allowing more than 99.9 \% of thermal neutrons to pass through to the reaction cell unattenuated. Backgrounds were measured with a beam block of $^6$Li glass positioned between the fission chamber and the reaction cell. This beam block absorbed nearly all thermal neutrons without attenuating gamma radiation incident on the cell. In this way, the gamma-ray and dark-current contributions to the PMT signal were measured. A diagram of the neutron beam setup appears in Figure \ref{Figure8}.  

\section{Model Calculations}

Monte Carlo techniques were employed to calculate both charged-particle transport from the $^{10}$B(n,$\upalpha$)$^7$Li  reaction and FUV photon emission within the reaction cell. The results of these calculations were used to determine the spatial distribution of charged particles escaping the boron films and the photon collection efficiency of the experimental apparatus.

The number of photons incident on the PMT depended upon the number of photons produced in the reaction cell, the spatial distribution of those emissions, the probability with which the photons were reflected from adjacent surfaces, and the type of reflection they underwent (i.e., specular or diffuse). Some photons intersected the PMT directly; others were reflected into the PMT from the surfaces surrounding the reaction volume; and a large majority were absorbed by the interior surfaces of the cylinder surrounding the $^{10}$B film without reaching the detector. The simplest model for the calculation of detector collection efficiency, $\epsilon$ assumes that all of the photons were emitted isotropically from a point source at the center of the $^{10}$B film surface, where
\begin{equation}
\epsilon = \frac{\Omega}{4\pi} = \frac{1}{2} \left( 1 - \frac{d}{\sqrt{d^2+a^2}} \right),
\end{equation}
$\Omega$ is the solid angle subtended by the detector, $d$ is the distance from the source to the detector photocathode (85.8 mm), and $a$ is the radius of the detector photocathode (11.5 mm). This approximation gives a collection efficiency of 0.443 \%. However, the approximation does not account for the reflectivities of the surfaces surrounding the reaction volume, the extended volume over which photon emission occurs, or variations in the spatial response of the PMT.

\begin{figure}[h]
\centering
\includegraphics{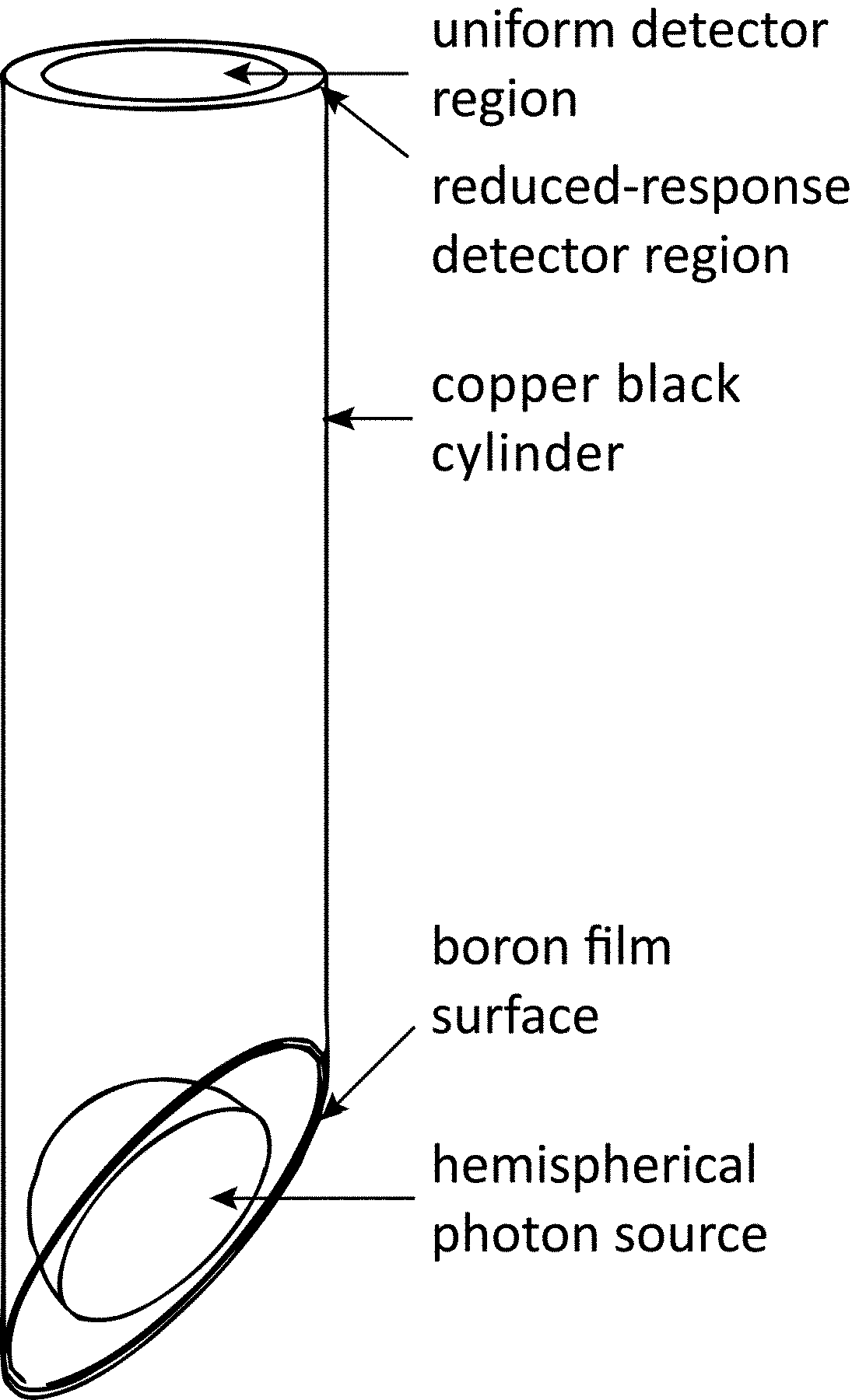}
\caption{The geometry of the photon collection model.}
\label{Figure9}
\end{figure}

A Monte Carlo routine was developed to incorporate the extended photon emission volume, the geometry of the reaction volume, the reflectivities of the various surfaces, and the measured position-dependent sensitivity of the PMT photocathode. The model geometry appears in Figure \ref{Figure9}. For each photon, an emission position was generated from a uniform and random distribution over the volume of a hemisphere tilted at 45$^{\circ}$ with respect to the detector plane. A direction for the photon path was then calculated from a uniform random distribution. Subsequently, the intersection of the path with one of the three surrounding surfaces (i.e., $^{10}$B film, black copper-oxide cylinder, and detector face) determined whether the photon was collected, reflected, or absorbed. 

Photons striking the $^{10}$B surface were specularly reflected with a reflection coefficient of 0.35, derived from measurements in \cite{27}. Photons striking the copper-oxide cylinder were diffusely reflected with a probability of 0.01, derived from measurements in \cite{28}. The final tally was weighted according to the data in Figure \ref{Figure6}: photons striking the uniform detector region were given a score of 1, and photons striking the outer region were given a score of 0.57. From $10^7$ particle histories, the collection efficiency, $\epsilon_\Omega$, was determined to be ($0.512 \pm 0.050$) \%. The statistical uncertainty in $\epsilon_\Omega$ was determined by calculating the standard deviation of 10 runs of $10^6$ histories each. The total uncertainty includes both the statistical uncertainty and systematic uncertainties in the model inputs (e.g., geometry, reflectivities).

The radius of the hemispherical photon emission volume in the photon collection model was approximated through modeling noble gas ionization density following the $^{10}$B(n,$\upalpha$)$^7$Li reaction with the radiation transport code, TRIM \cite{29}. The TRIM input files specified the thicknesses of the boron films and the noble gas pressures, as well as the emission positions, directions, and starting energies of the $^{10}$B(n,$\upalpha$)$^7$Li reaction products. The starting positions of the reaction products within the $^{10}$B film were sampled using an inverse transform for the depth profile and the neutron beam width.

\begin{figure*}[t]
\centering
\includegraphics{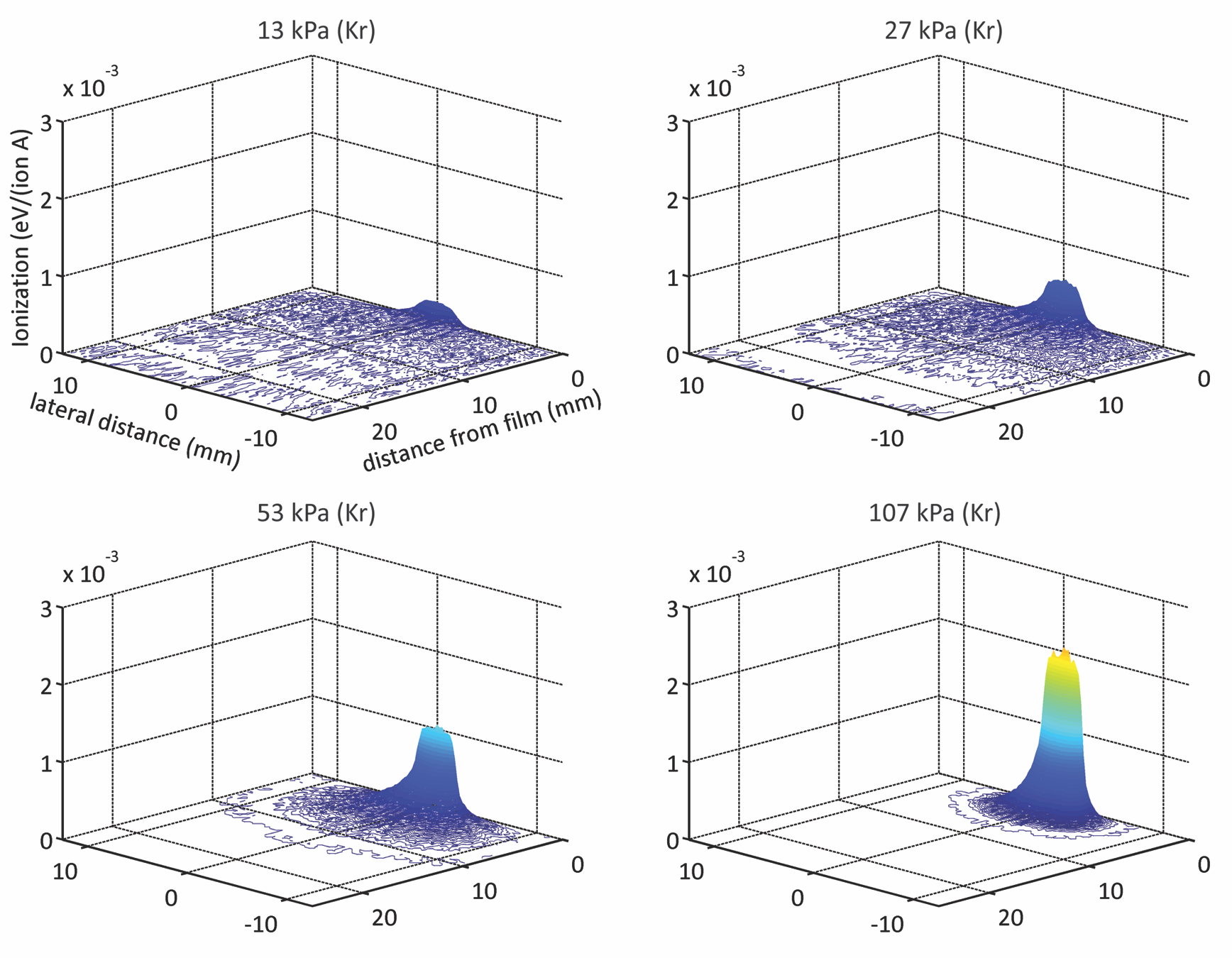}
\caption{Ionization density in Kr at various pressures from neutron irradiation of a 300 nm $^{10}$B film. Each right-hand, upright plane represents the surface of the film.}
\label{Figure10}
\end{figure*}

Ionization density distributions from $^{10}$B(n,$\upalpha$)$^7$Li in a 300 nm $^{10}$B film under various pressures of Kr appear in Figure \ref{Figure10}. The plots show the two-dimensional shape (x-y plane) and the ionization density (z axis) of the volume in which the charged particles deposited their energy. Contours are drawn at increments of $3 \times 10^{-6}$ eV/$\mathrm{\AA}$ per ion. Based on the results of these simulations, a radius of 10 mm was chosen for the hemispherical photon emission volume in the photon collection model. However, as shown in Figure \ref{Figure10}, the size of the emission volume was pressure dependent and the shape was non-uniform. 

In summary, the photon collection model assumed the following: (1) a hemispherical  excimer photon emission volume, (2) a uniform, pressure-independent distribution of NGE emissions within the volume, (3) accurate values of the reflectivities of boron and copper black, (4) negligible contributions of secondary and tertiary reflections, and (5) the absence of refraction by the MgF$_2$ windows.

\section{Results and Discussion}

\subsection{Photon detector package efficiency}

The photon detector package (i.e., MgF$_2$ windows, evacuated volume, and PMT) was calibrated as a unit at SURF III, beamline 4 (BL-4). The efficiency measurements accounted for absorption by the external and PMT MgF$_2$ windows, the quantum efficiency of the PMT, and the thresholds of the counting electronics. A silicon photodiode that was previously calibrated against a cryogenic radiometer \cite{30,31} was used to obtain the absolute response of the photon detector package. The calibration covered the FUV region common to the NGE spectra, between 130 nm and 210 nm. The PMT was operated in the pulse counting mode with electronics identical to those used in the scintillation yield measurements.

The detector calibration was performed in several stages due to the large difference in sensitivity of the photodiode and the PMT. The incident photon flux was measured with the photodiode over the wavelength range 130 nm to 210 nm in 5 nm increments, at electron beam currents of approximately (50, 40, 30, 20, and 10) mA. These measurements were used to determine a linear relationship between electron beam current in the SURF storage ring and photon flux in the detector position, at each discrete wavelength. 

Following these measurements, the SURF electron beam current was reduced to (10, 5, and 1) $\upmu$A. At these lower beam currents, the photon flux was within the operating range of the PMT. Spectral scans over the same wavelength region were repeated at these reduced currents. The intrinsic efficiency of the photon detector package, $\epsilon_i$, was determined by dividing the number of pulses observed from the PMT at each wavelength by the number of photons incident on the photon detector package at that wavelength. The results appear as discrete points in Figure \ref{Figure5} for three electron beam currents. The relative uncertainties in $\epsilon_i$ include both statistical uncertainties and uncertainties in the linear regressions.

Because NGE emissions occur over broad continua, an effective photon detection efficiency was determined for the three noble gases used in the NGE scintillation experiments. The effective efficiency, $\hat{\epsilon}_i$, was calculated with a continuous weighted average, in the form,
\begin{equation}
\hat{\epsilon}_i = \frac{\int \epsilon_i(\lambda) s(\lambda)\mathrm{d}\lambda}{\int s(\lambda)\mathrm{d}\lambda},
\end{equation}
where $\epsilon_i(\lambda)$ is a fit of the discrete values of $\epsilon_i$, and $s(\lambda)$ is the wavelength distribution of the NGE continua obtained by digitizing excimer emission spectra from \cite{1}. The values of $\hat{\epsilon}_i$ for Ar, Kr, and Xe were 1.65 \%, 3.14 \%, and 2.61 \%, respectively. The uncertainties in $\hat{\epsilon}_i$ were derived from the average value of the relative uncertainties in $\epsilon_i$. 

\subsection{Areal densities of the $^{10}$B thin films}

The thermal-neutron absorption properties of the $^{10}$B thin films were measured at the NIF. Thermal neutrons from the NCNR reactor passed through a cooled, single-crystal bismuth filter, a series of apertures, and an evacuated flight tube before impinging on the samples with a fluence of $5.31 \times 10^6$ cm$^2$s$^{-1}$. The NIF detector consists of a $^6$Li conversion layer, a ZnS scintillation layer, and an x-ray imager made of amorphous silicon.

Each boron film used in the scintillation experiments was placed in a holder and mounted at the imaging station. The neutron beam illuminated each sample, and the detector collected a series of 1800 images with one-second exposures. This measurement was repeated to obtain a flat-field image without a sample or a sample holder in the beam. All of the images were then corrected for the point spread function (PSF) of the detector system. The PSF is a systematic additive background attributed to diffuse light in the scintillation screen of the detector.

The images of each sample were then averaged to form a single image. By dividing the averaged image of each boron film by the averaged flat-field image, the fractional absorption of each sample was determined. The fractional absorption of each pixel, $P$, in the resulting images has the form,
\begin{equation}
P = 1 - \frac{I}{I_0} = 1 - e^{-\Sigma x},
\end{equation}
where $I_0$ is the intensity of the incident neutron beam, $I$ is the intensity of the neutron beam after passing through the sample, $\Sigma$ is the macroscopic neutron absorption cross section of the sample, and $x$ is the thickness of the sample. The value of $I/I_0$ was determined by dividing each pixel value in the sample image by the corresponding pixel value in the empty flat-field image. 

By rearranging Equation 3 to solve for $\Sigma x$ and averaging the values of $\Sigma x$ over the surface of each film, the areal density of each film, $\rho_{A1}$, was determined with the equation,
\begin{equation}
\rho_{A1} = \frac{\overline{(\Sigma x)} M_1}{\hat{\sigma}_{\mathrm{NIF}}N_A},
\end{equation}
where $\overline{(\Sigma x)}$ is the average value of $\Sigma x$, $\hat{\sigma}_{\mathrm{NIF}}$ is the effective neutron absorption cross section of $^{10}$B in the NIF neutron beam,  $M_1$ is the molar mass of $^{10}$B, and $N_A$ is Avogadro's number. The value of $\hat{\sigma}_{\mathrm{NIF}}$  was determined with a continuous weighted average, in the form,
\begin{equation}
\hat{\sigma}_{\mathrm{NIF}} = \frac{\int \sigma_1(\lambda) \phi(\lambda)\mathrm{d}\lambda}{\int \phi(\lambda)\mathrm{d}\lambda},
\end{equation}
where $\sigma_1(\lambda)$ is the wavelength-dependent, microscopic absorption cross section of $^{10}$B, and $\phi(\lambda)$ is the neutron wavelength distribution of the NIF beam. This neutron wavelength distribution was previously measured at the NIF with a neutron chopper and time-of-flight spectrometry. The thermal neutron absorption cross section of silicon is small (2.16 b \cite{24}); thus, the absorption of each sample was attributed completely to absorption by $^{10}$B. From Equation 4, the resulting areal densities of the $^{10}$B thin films were (73.5, 140, 209, and 260) $\upmu$g/cm$^2$.

A conservative 10 \% uncertainty was assigned to the values of $\rho_{A1}$. This uncertainty arises predominantly from the uncertainty in $\hat{\sigma}_{\mathrm{NIF}}$. However, the relative root mean square (RMS) variations in $\overline{(\Sigma x)}$̅  were calculated to be $< 2$ \%. The RMS values were nearly identical for all of the measured samples, indicating a statistical limit imposed by counting time or baseline noise of the NIF detector, rather than density variations in the $^{10}$B samples. The RMS values do, however, provide an upper limit for this density variation.

\subsection{Neutron beam fluence and absorption}

The neutron beam from the MUTR thermal column was characterized to determine the neutron absorption rate in the $^{10}$B films. During the NGE scintillation measurements, a NIST-calibrated $^{235}$U fission chamber was used to continuously monitor the fluence of the thermal-column neutron beam. By combining these measurements with the areal densities of the $^{10}$B thin films, the neutron absorption rate of the $^{10}$B films was determined.

The neutron fluence incident on the boron film, $N_b$, was derived from fission chamber measurements, $n_n$, with the equation,
\begin{equation}
N_b = \frac{n_n}{t_n}\frac{M_2 \zeta}{\hat{\sigma}_2 \rho_{A2} N_A},
\end{equation}
where $t_n$ is the length of each neutron count period, $M_2$ is the molar mass of $^{235}$U, $\hat{\sigma}_2$ is the effective microscopic absorption cross section of $^{235}$U, $\rho_{A2}$ is the areal density of the fission chamber deposit (($458.1 \pm 2.3$) $\upmu$g/cm$^2$), and $\zeta$ is a self-absorption correction factor (1.038). The areal density of the fission chamber deposit was previously measured at NIST. The factor $\zeta$ accounts for self-absorption of fission fragments in the $^{235}$U deposit. 

The rate at which neutrons were absorbed by the boron films, $N_n$, was derived with the equation,
\begin{equation}
N_n = N_b \frac{\hat{\sigma}_1 \rho_{A1} N_A}{M_1} \mu \sqrt{2},
\end{equation}
where $\hat{\sigma}_1$ is the effective microscopic absorption cross section of $^{10}$B, and $\mu$ is the fraction of the neutron beam that was not absorbed by material between the fission chamber and the boron target (0.995). The factor $\sqrt{2}$ was included to account for the 45$^{\circ}$ tilt of the film with respect to the incident neutron beam. 

Due to the need for sufficient counting statistics, neutron count periods often overlapped multiple photon count periods. To determine a value of $n_n$ corresponding to each photon count, a fit was performed on the neutron count measurements from each day, $\hat{n}_n$. The standard deviation in the counts from each day provided the uncertainty in the fitted values.

The remaining unknowns in Equations 6 and 7 are the effective microscopic cross sections, $\hat{\sigma}_1$ and $\hat{\sigma}_2$. Equation 5 describes an effective microscopic cross section, and this equation may be rewritten in the form,
\begin{equation}
\hat{\sigma} = \frac{\int \sigma(v) n(v) v \mathrm{d}v}{\int n(v) v \mathrm{d}v},
\end{equation}
where $\sigma(v)$ is the velocity-dependent microscopic cross section, $n(v)$ is the velocity-dependent neutron population, and $v$ is the neutron velocity. In the $1/v$ absorption region of $\sigma(v)$, $\sigma \propto v^{-1}$; therefore, the product $\sigma v$ is a constant. Accordingly, $\sigma v$ may be pulled out of the integral and the ratio of two effective cross sections may be reduced to a constant \cite{32}. Since $^{10}$B and $^{235}$U are nearly ideal $1/v$ absorbers in the thermal region, this approximation may be applied to the effective cross sections in Equation 7, which become the constant, $\kappa$. This reduction holds true as long as contributions from neutrons outside the $1/v$ energy region are negligible, which was previously demonstrated through MCNP modeling of the MUTR \cite{33} and direct measurement \cite{34}. While $\kappa$ is not perfectly constant below 1 eV, it varies by only 2.6 \% between 0.0125 eV and 0.0375 eV. The value of $\kappa$ at 0.025 eV ($6.57 \pm 0.17$) was selected for the yield calculations. 

Ultimately, Equation 7 becomes,
\begin{equation}
N_n = \frac{\hat{n}_n}{t_n} \frac{\rho_{A1} M_2}{\rho_{A2} M_1} \kappa \mu \zeta \sqrt{2}.
\end{equation}
Depending on the $^{10}$B film thickness, the average neutron absorption rate ranged from 90 s$^{-1}$ to 330 s$^{-1}$.

\subsection{Yield calculation, results, and uncertainties}

The NGE scintillation yield from the $^{10}$B(n,$\upalpha$)$^7$Li reaction was measured as a function of gas type, gas pressure, and $^{10}$B target thickness. The scintillation yield per neutron absorption, $Y$, was calculated with the equation,
\begin{equation}
Y = \frac{N_{h \nu}}{N_n},
\end{equation}
where $N_{h \nu}$ is the rate at which excimer photons were generated in the reaction cell, and $N_n$ is the rate at which neutrons were absorbed in the boron targets (as shown in Equation 9). Here,
\begin{equation}
N_{h \nu} = \frac{n_{h \nu}}{\epsilon_\Omega \hat{\epsilon}_i t_{h \nu}},
\end{equation}
where $n_{h \nu}$ is the number of signal pulses generated by the PMT over time $t_{h \nu}$, $\epsilon_\Omega$ is the photon collection efficiency, and $\hat{\epsilon}_i$ is the intrinsic efficiency of the photon detector package. 

The number of pulses generated by the PMT, $n_{h \nu}$, was calculated from two consecutive measurements. The first measurement included both the NGE signal and background. The second measurement was gathered with the lithium-glass beam block in front of the reaction cell and included background contributions from dark current, gamma-ray interactions with the PMT and the solid components of the reaction cell, and gamma-ray interactions with the noble gases in the reaction cell. The first two background sources produced a baseline count rate independent of the gas pressure in the cell. The third source was dependent on noble gas type and pressure. The final yield measurements account for the statistical uncertainties in $n_{h \nu}$.

\begin{figure}[!h]
\centering
\includegraphics{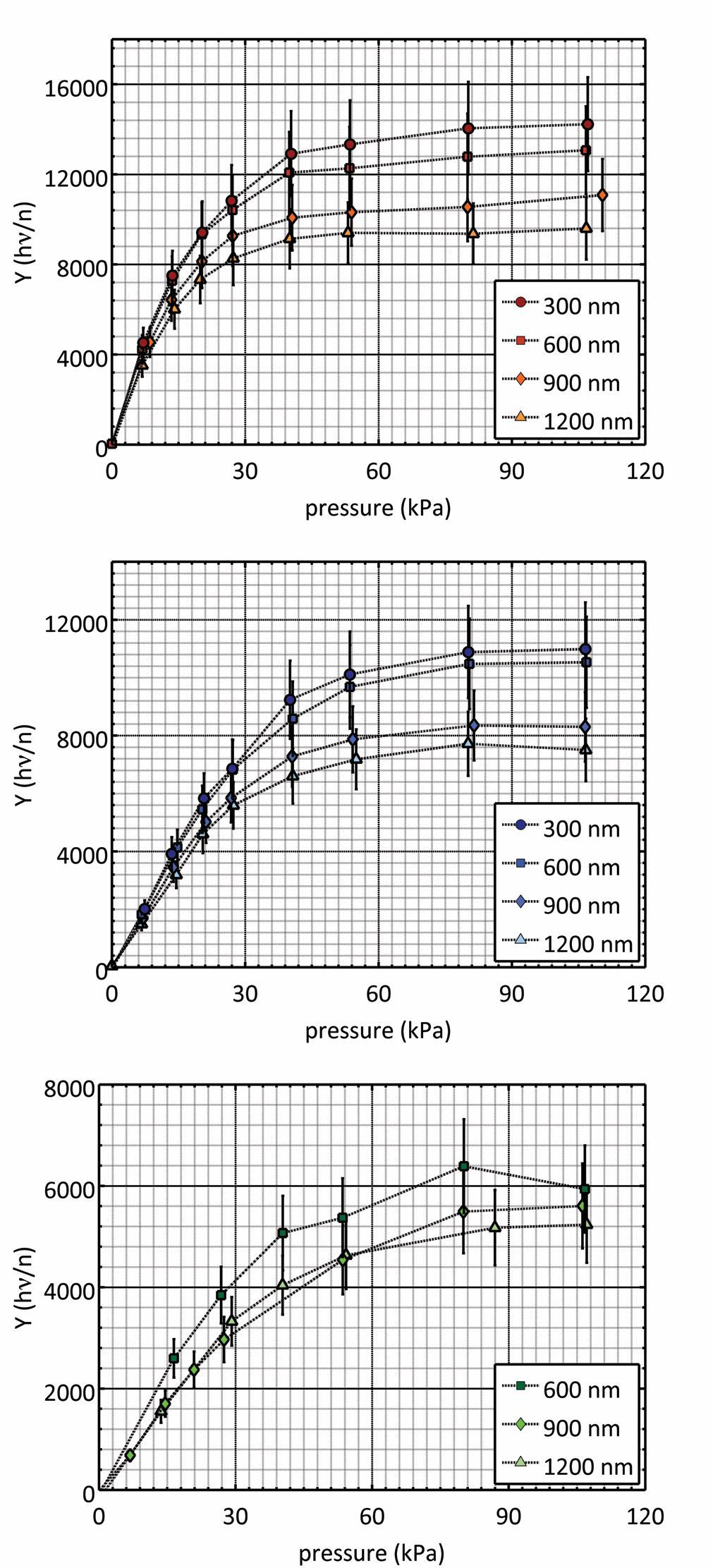}
\caption{Excimer scintillation yield from $^{10}$B(n,$\upalpha$)$^7$Li in Xe (top), Kr (middle), and Ar (bottom) at pressures to 107 kPa. Y-axis values represent excimer photons per neutron absorbed. Insets distinguish different $^{10}$B target thicknesses. Error bars represent combined uncertainties from various inputs as described in Table \ref{Table1}.}
\label{Figure11}
\end{figure}

Plots of $Y$ appear in Figure \ref{Figure11}. The results show yields of (5200 to 6000) photons per neutron absorption for Ar, (7500 to 11,000) photons per neutron absorption for Kr, and (9600 to 14,200) photons per neutron absorption for Xe at pressures of 107 kPa. 

The total uncertainty in $Y$ was calculated by propagating the uncertainties of each of its factors listed in Table \ref{Table1}. The resulting relative standard uncertainties in $Y$ range between 14.7 \% and 15.5 \%, and are reflected by the error bars in Figure \ref{Figure11}.

\begin{table*}[t]
\caption{Values and standard relative uncertainties ($\delta$) of the inputs to the yield calculation.}
\label{Table1}
\begin{tabular}{l l l l l l}
\hline \hline

& Quantity & Description & Value (Unit) & $\delta$ (\%) & Source \\ \hline

\multirow{4}{*}{$N_{h\nu}$} & $n_{h \nu}$ & excimer photon counts & 1000--80,000 & 0.5--5 & scintillation measurements \\

& $\hat{\epsilon}_i$ & detector efficiency & 1.6--3.1 (\%) & 3.5 & detector calibration; excimer spectra \\

& $\epsilon_\Omega$ & collection efficiency & 0.512 (\%) & 9.7 & photon collection modeling \\

& $t_{h\nu}$ & photon count time & 200 (s) & -- & controlled \\ \hline
	
\multirow{9}{*}{$N_n$} & $\hat{n}_n$ & neutron counts (fitted) & 2730--3150 & 1.6--4.4 & fission chamber measurements \\

& $\kappa$ & thermal cross section ratio & 6.57 & 2.6 & derived from tabulated values \\

& $\rho_{A1}$ & areal density of $^{10}$B films & 73.5--260 ($\upmu$g/cm$^2$) & 10 & neutron imaging \\

& $\rho_{A2}$ & areal density of $^{235}$U deposit & 458.1 ($\upmu$g/cm$^2$) & 0.5 & NIST measurement \\

& $M_1$ & molar mass of $^{10}$B & 10.0129 (g/mol) & -- & tabulated \\

& $M_2$ & molar mass of $^{235}$U & 235.0439 (g/mol) & -- & tabulated \\

& $\mu$ & transmission fraction & 0.995 & -- & transmission calculation \\

& $\zeta$ & self-absorption factor & 1.038 & -- & previous NIST calculation \\

& $t_n$ & neutron count time & 1200 (s) & -- & controlled \\

\hline \hline
\end{tabular}
\end{table*}

\section{Analysis}

In the experiments described above, the $^{10}$B(n,$\upalpha$)$^7$Li reaction generated large excimer scintillation signals as a result of charged-particle reaction products transferring their energy through electronic excitation and ionization of noble-gas atoms. NGE yields increased with the atomic number of the noble gases as excimer excitation energies declined from 8.9 eV for Ar, to 8.0 eV for Kr, to 7.0 eV for Xe \cite{35}. Photon yields also increased with gas pressure as a result of an increased frequency of collisions between charged particles and noble gas atoms, an increased frequency of excimer-forming collisions between excited and ground state noble gas atoms, and higher photon collection efficiency for smaller photon-emission regions.  

At 80 kPa for Ar, 53 kPa for Kr, and 40 kPa for Xe, NGE emissions reached a plateau, indicating that at these gas pressures no charged particle energy was lost to the cell walls, and the photon emission volumes were produced completely within the line-of-sight of the photon detector. This conclusion was confirmed by the results of the TRIM simulations discussed above. Additionally, NGE yields decreased with increasing target thickness due to the larger average energy loss of charged particles escaping thicker films. The saturation pressures were dependent on the geometry of our reaction cell (i.e., the radius of the photocathode and the distance of the chamber walls from the location of charged-particle emission).  In the saturation region, when the photon emission volume was small and well defined, photon yields depended only on the properties of the neutron absorbing targets and not on the remaining detector geometry.

To understand the excimer photon yield per captured neutron, excimer production and decay must be considered. As energetic charged particles travel through noble gases, they deposit their energy by excitation and ionization of the surrounding noble gas atoms. These processes include,
\begin{eqnarray}
R + A & \to& e + A^+ + R', \\
R + A & \to& A^* + R', \\
e + A^+ & \to& A^*,
\end{eqnarray}
where $R$ is a charged particle, $A$ is a noble gas atom, $e$ is an electron, $A^+$ is an ionized noble gas atom, $R'$ is a charged particle of reduced energy, and $A^*$ is a noble gas atom in an excited electronic state \cite{35}. Excited noble-gas atoms may decay through the radiative emission of a FUV photon ($h\nu_a$),
\begin{equation}
A^* \to A + h\nu_a.
\end{equation}

At increased pressures (103 Pa to 104 Pa), the probability of another atom reabsorbing this atomic photon increases, and the photons become ``trapped” within the gas volume. Independently, the likelihood of three-body collisions also increases at higher pressures, leading to excimer formation,
\begin{equation}
A^* + 2A \to A_2^* + A,
\end{equation}
where $A_2^*$ is a NGE. This reaction occurs on a time scale of $10^{-11}$ s to $10^{-12}$ s \cite{35}. The excimers may then decay by emitting a photon,
\begin{equation}
A_2^* \to 2A + h\nu_m.
\end{equation}

Variations in the amount of energy carried away by the third body, as in Equation 16, and the repulsive nature of the molecular ground state enable NGE emissions to occur over broad molecular continua. Spontaneous radiative decay times for simple excimers fall between 0.005 $\upmu$s and 5 $\upmu$s \cite{36}. Comparatively, dissociation times for weakly bound molecular ground states may be as short as $10^{-13}$ s, on the timescale of a single molecular vibration. This range of decay times is attributed to the different transition probabilities between the excimer singlet and triplet states and the singlet ground state. The forbidden triplet to singlet-state transition may have a characteristic decay time two orders of magnitude longer than the singlet to singlet-state decay. Collisions play an important role in the decay process, thus decay times are strongly dependent on gas pressure \cite{3,36}. At higher pressures, triplet-singlet transitions, aided by electronic and atomic collisions, occur with increased probability. Therefore, both the formation and the decay of excimers are more rapid at high gas pressures. 

\section{Conclusions}

The observed excimer scintillation signals from the $^{10}$B(n,$\upalpha$)$^7$Li reaction are comparable to the yields of some liquid and solid neutron scintillators (for yields of other scintillators see \cite{37}). The low densities of noble gases in these experiments translate to gamma sensitivities that are inherently lower than liquid and solid scintillators. We attribute the increase in the scintillation yield with noble gas pressure to the increased frequency of interactions between charged-particle reaction products and noble gas atoms, and the increased probability of three-body excimer-forming collisions. 

The results of this work indicate the potential for NGE to provide an efficient means of neutron detection. The results also suggest ways in which to modify the current arrangement to greatly increase detection efficiency. The $^{10}$B films used in these experiments absorbed at most 8 \% of the incident neutrons. Thicker films will lead to increased absorption, but will also attenuate the average energy of the emerging ions. To avoid this attenuation, $^{10}$B films may be stacked in an array. While absorption in any single film will be small, the total absorption through an array of 30 $^{10}$B films, 1 $\upmu$m thick, for example, approaches 80 \%. In the present experiment, ions emitted toward the silicon substrate are often absorbed by it. As a result, the ion yield is reduced by approximately 50 \%. Supporting the $^{10}$B films on structures that allow the unimpeded passage of the product ions can increase efficiency.

A very large increase in efficiency can also be attained by increasing the solid angle subtended by the photon detector, the quantum efficiency of the photon detector, and the reflectivity of the surfaces in the reaction cell. Increasing the solid angle may be accomplished with large area detectors placed close to the $^{10}$B film target. While this is not practical with existing FUV PMTs, rapid advances in extending the short wavelength sensitivity of large silicon photomultipliers promise to produce practical devices that can be used to detect the FUV photons with high efficiency over a large solid angle \cite{20,38}.  Additionally, the reflectivity and shapes of the reaction cell surfaces may be further optimized for efficient photon collection.

\section*{Acknowledgements}

The authors would like to thank Patrick Hughes, Daniel Hussey, David Jacobsen, David Gilliam, Craig Heimbach, Jeffrey Nico, Pieter Mumm, and Muhammad Arif from the NCNR; Uwe Arp, Alex Farrell, Edward Hagley, Thomas Lucatorto, Mitchell Furst, Ping Shaw, and Steve Grantham from SURF; Gerard Henein from the CNST; John Abrahams and Thomas Loughran from the Maryland NanoCenter; Karen Gaskell from the Maryland Materials Research Science and Engineering Center; June Tveekrem from NASA Goddard Space Flight Center; and Lis Nanver from Delft University of Technology.

The authors would like to thank the U.S. Nuclear Regulatory Commission (NRC) for its financial support (NRC fellowship grant NRC-38-09-934).


\begin{thebibliography}{38}

\bibitem{1}
R. E. Huffman, J. C. Larrabee, and Y. Tanaka, Appl. Optics 4, 1581 (1965).
    
\bibitem{2}
T. E. Stewart, G. S. Hurst, T. E. Bortner, J. E. Parks, F. W. Martin, and H. L. Weidner, J. Opt. Soc. Am. 60, 1290(1970).
    
\bibitem{3}
M. Mutterer, P. Grimm, H. Heckwolf, J. Pannicke, W. Spreng, and J. Theobald, Lect. Notes Phys. 178, 63 (1983).

\bibitem{4}
J. J. Hopfield, Astrophys, J. 72, 133 (1930).

\bibitem{5}
G. Herzberg, \emph{Molecular Spectra and Molecular Structure I. Spectra of Diatomic Molecules, 2nd edition}, (D. van Nostrand, Princeton, 1950) p. 536.

\bibitem{6}
Y. Tanaka, J. Opt. Soc Am. 45, 710 (1955).

\bibitem{7}
Y. Tanaka, A. S, Jursa, and F. J. LeBlanc, J. Opt. Soc. Am. 48, 304 (1958).

\bibitem{8}
B. Eliasson and U. Kogelschatz, Appl. Phys. B 46, 299 (1988).

\bibitem{9}
M. Mutterer, J. P. Theobald, and K.-P. Schelhaas, Nuc. Instrum. Methods 144, 159 (1977).

\bibitem{10}
C. A. N. Conde, A. J. P. L. Policarpo, Nuc. Instrum. Methods 53, 7 (1967).

\bibitem{11}
T. Takahashi, S. Himi, M. Suzuki, J-Z. Ruan, and S. Kubota, Nuc. Instrum. Methods 205, 591 (1983).

\bibitem{12}
R. A. Nobles, Rev. Sci. Inst. 27, 280 (1956).

\bibitem{13}
A. J. P. L. Policarpo, Phys. Scripta, 23, 539 (1981).

\bibitem{14}
P. Grimm, F.-J. Hambsch, M. Mutterer, J. P. Theobald, and S. Kubota, Nuc. Instrum. Meth. A 262, 394 (1987).

\bibitem{15}
A. Morozov, T. Heindlkm, R. Krücken, A. Ulrich, and J. Wieser, J. Appl. Phys. 103, 103301 (2008).

\bibitem{16}
J. B. Birks, \emph{Excimers}, Rep. Prog. Phys. 38, 903 (1975).

\bibitem{17}
S. Kubota, T. Takahashi, and T. Doke, Phys. Rev. 165, 225 (1968).

\bibitem{18}
D. C. Lorents, Rad. Res. 59, 438 (1974).

\bibitem{19}
E. Aprile and T. Doke, Rev. Mod. Phys. 82, 2053 (2010).

\bibitem{20}
R. Chandrasekharan, M. Messina, and A. Rubbia, Nucl. Instrum. Meth. A 546, 426 (2005).

\bibitem{21}
K. Saito, H. Tawara, T. Sanami, E. Shibamur, and S. Sasaki, IEEE T. Nucl. Sci. 49, 1674 (2002).

\bibitem{22}
E. Morikawa, R. Reininger, P. Gurtler, and V. Saile, J. Chem. Phys. 91, 1469 (1989).

\bibitem{23}
P. P. Hughes, M. A. Coplan, A. K. Thompson, R. E. Vest, and C. W. Clark, Appl. Phys. Lett. 97, 234105 (2010).

\bibitem{24}
G. Hale and P. Young, ``ENDF/B-VII.” http://www.nndc.bnl.gov/exfor/endf00.jsp, April 2006.

\bibitem{25}
D. S. Hussey, D. L. Jacobson, M. Arif, P. R. Huffman, R. E. Williams, and J. C. Cook, Nucl. Instrum. Meth. A 542, 9 (2005).

\bibitem{26}
U. Arp, C. Clark, L. Deng, N. Faradzhev, A. Farrell, M. Furst, S. Grantham, E. Hagley, S. Hill, T. Lucatorto, P.-S. Shaw, C. Tarrio, and R. Vest, Nucl. Instrum. Meth. A 649, 12 (2011).

\bibitem{27}
M. Vidal-Dasilva, M. Fernandez-Perea, J. A. Mendez, J. A. Aznarez, and J. I. Larruquert, Appl. Optics 47, 2926 (2008).

\bibitem{28}
M. P. Newell and R. A. M. Keski-Kuha, Appl. Optics 36, 5471 (1997).

\bibitem{29}
J. F. Ziegler, ``SRIM---The Stopping Range of Ions in Matter.” 2010. 

\bibitem{30}
C. R. Yokley in \emph{Applications of Optical Metrology--Techniques and Measurements II}, Bellingham, Washington, 7–8 April 1983, edited by J. J. Lee, SPIE, pp. 2--8.

\bibitem{31}
T. J. Quinn and J. E. Martin, Philos. Trans. Roy. Soc. London A 316, 85 (1985).

\bibitem{32}
K. S. Krane, \emph{Introductory Nuclear Physics} (Wiley \& Sons, USA, 1988) pp. 451--453.

\bibitem{33}
A. B. Mohamed, Ph.D. thesis, University of Maryland, College Park, 2009.

\bibitem{34}
I. Gifford, Ph.D. thesis, University of Maryland, College Park, 2013.

\bibitem{35}
E. Aprile, A. Bolotnikov, A. Bolozdynya, and T. Doke, \emph{Noble Gas Detectors}, (Wiley VCH, Berlin 2006).

\bibitem{36}
M. H. R. Hutchinson, Appl. Phys. 21, 95 (1980).

\bibitem{37}
C. W. E. van Eijk, A. Bessiere, and P. Dorenbos, Nucl. Instrum. Meth. A 529, 260 (2004).

\bibitem{38}
L. M. P. Fernandes, F. D. Amaro, A. Antognini, J. M. R. Cardoso, C. A. N. Conde, O. Huot, P. E. Knowles, F. Kottmann, J. A. M. Lopes, L. Ludhova, C. M. B. Monteiro, F. Mulhauser, R. Pohl, J. M. F. dos Santos, L. A. Schaller, D. Taqqu, and J. F. C. A. Veloso, JINST 2, P08005 (2007).

\end{thebibliography}
\end{document}